\documentstyle[epsfig] {aipproc}
\begin{document}

\title{ The Lick Observatory Supernova Search}

\author{W. D. Li, A. V. Filippenko, R. R. Treffers, A. Friedman,\\ 
E. Halderson, R. A. Johnson, J. Y. King, M. Modjaz, M. Papenkova,\\
Y. Sato, and T. Shefler}
\address{ Department of Astronomy,
University of California, Berkeley, CA 94720-3411 USA\\
email: (wli, alex, rtreffers)@astro.berkeley.edu
}

\maketitle

\begin{abstract}
We report here the current status of the Lick Observatory Supernova
Search (LOSS) with the Katzman Automatic Imaging Telescope (KAIT).
The progress on both the hardware and the software of the system
is described, and we present a list of recent discoveries. LOSS is
the world's most successful search engine for nearby supernovae.
\end{abstract}

\section*{ Introduction}

Located at Lick Observatory atop Mount Hamilton east of San Jose,
California, the 0.75-m Katzman Automatic Imaging Telescope
(KAIT) is a robotic telescope dedicated to the Lick Observatory
Supernova Search (LOSS) and the monitoring of variable celestial 
objects. It is equipped with a CCD camera and an automatic autoguider 
(that is, the autoguider is able to find its own guide stars).

KAIT is the third robotic telescope in the Berkeley Automatic 
Imaging Telescope (BAIT) program. The predecessors to KAIT were two 
telescopes developed at the Leuschner Observatory, which is located
about 10 miles east of the campus of the University of California,
Berkeley. KAIT inherits the operational concept and the majority 
of the software from its two predecessors. More thorough descriptions
of the BAIT system can be found in references 1--4.

LOSS discovered its first supernova in 1997 (SN 1997bs in NGC 3627; 
Treffers et al. 1997 [5]). Its performance improved 
dramatically in 1998 and 19 supernovae (SNe) were discovered. In 
1999, 35 SNe were discovered by mid-December.

Multicolor photometry of SNe is an important scientific goal of 
KAIT. Because of the early discoveries of most of the LOSS SNe, many 
good light curves have been obtained.

We report our hardware and software setups for LOSS in Section 2,
the SN search in Section 3, and the discoveries and follow-up 
observations in Section 4.

\section*{ The Hardware and Software of LOSS}

\begin{figure}[b!]
\centerline{\epsfig{file=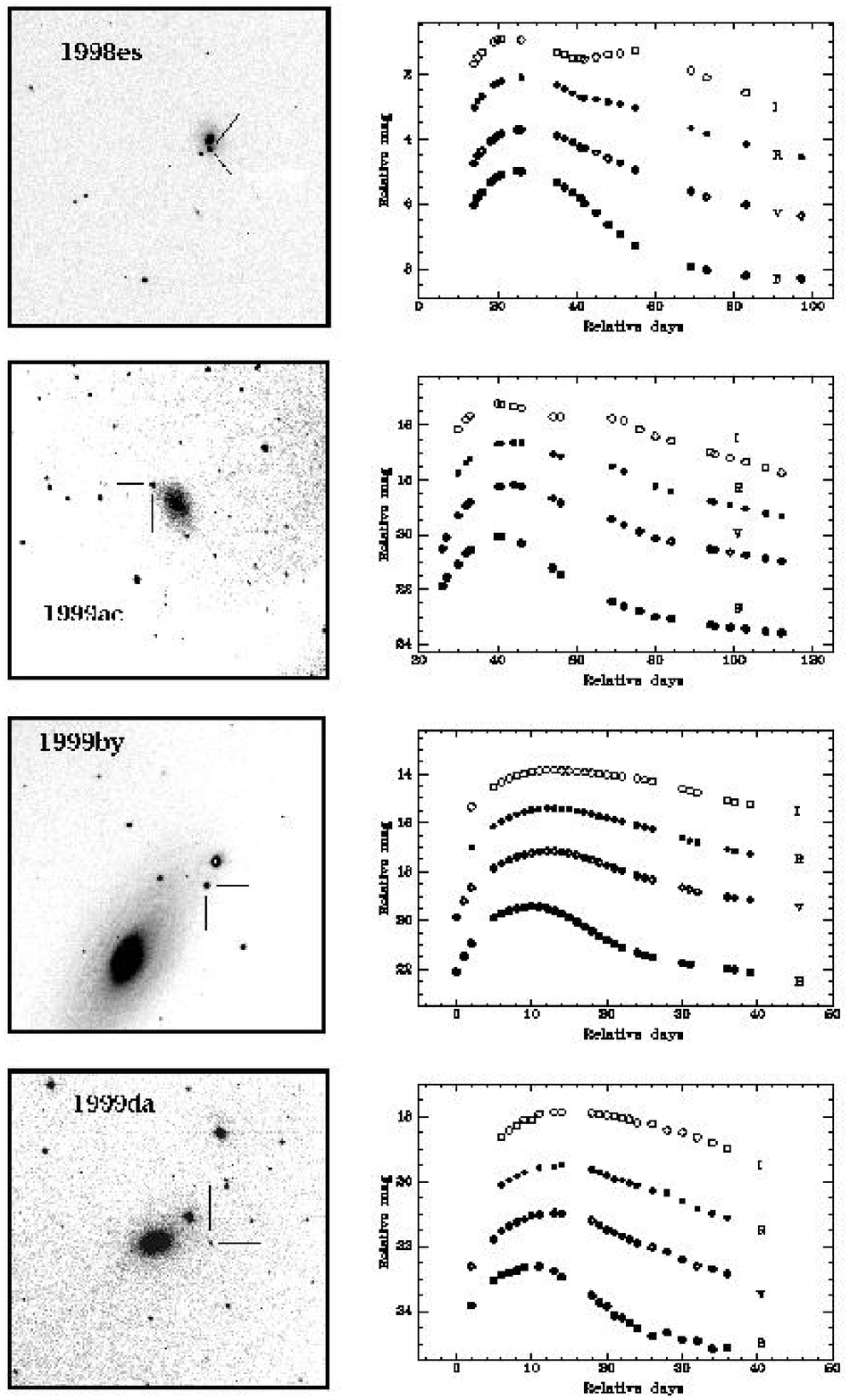,height=7.5in,width=5.7in}}
\vspace{10pt}
\caption{Some of the LOSS SN discoveries and their light curves. 
Examples shown (top to bottom) are the type Ia SNe 1998es, 1999ac, 
1999by, and 1999da. The light curves have been arbitrarily shifted 
up and down for display purposes.}
\label{fig1}
\end{figure}

KAIT has a 30-inch diameter primary with a Ritchey-Chreti\'en mirror
set. The focal ratio is $f/8.2$ which results in a plate scale of 
$33.2^{\prime\prime}$ mm$^{-1}$ at the focal plane. The telescope has 
a very compact design; it is lightweight and slews fast. An off-axis 
guider designed by one of us (RRT) enables the telescope to obtain 
long exposures.

The CCD camera is an Apogee AP7 with a SITe 512$\times$512 pixel 
back-illuminated chip. It is thermoelectrically cooled to about 
$60^\circ$C below the ambient temperature. The quantum efficiency 
(QE) is good (peak 60\%) and flat from 3000~\AA\ to 8000~\AA. The 
field of view is $6.7^{\prime}\times6.7^{\prime}$ with a scale of 
$0.8^{\prime\prime}$ pixel$^{-1}$. 

Observations done by KAIT are fully robotic. All hardware (telescope, 
filters, autoguider, CCD camera, slit, dome, weather station, etc.) are 
automatically controlled by the software (see Richmond, Treffers, and
Filippenko 1993 for details). Dark, bias, and twilight flatfield 
observations are also done automatically. A focusing routine finds a 
good focus for the telescope in twilight, then runs every 90 minutes 
during the night. The images are automatically transferred to the U.C. 
Berkeley campus to be processed. The appropriate template image is
subtracted from each galaxy image, and new objects are detected in the 
resulting images. The most promising candidates are reobserved
during the same night, while others require human evaluation before
rescheduling.

\section*{ The Supernova Search}

The LOSS galaxy sample includes about 5,000 nearby galaxies. Mosaic 
images are taken for some large, nearby galaxies. An automatic scheduler 
selects targets to be observed during the night according to their 
observation history. Follow-up observations of SNe or routine monitoring 
of some other objects (active galactic nuclei, variable stars, etc.) are 
also scheduled at the same time.

We have optimized the system in every possible way to increase the 
observation efficiency. The search images are taken through a hole 
in the filter wheel (i.e., no filter is used). This greatly 
increases the observation efficiency compared to observations through 
an $R$-band filter. The exposure time for the search images is only 25 
seconds, but because of the high QE of the CCD camera we still reach 
a limiting magnitude of $\sim 19$ (sometimes deeper). The order of 
galaxies to be observed is optimized so as to minimize the accumulated
movement of the telescope and dome during the night. Currently the 
observing efficiency is about 75 images per hour. KAIT can obtain more 
than 1,000 images during a winter night. 

Because of our high observing efficiency, all the sample galaxies
are observed every 3 to 5 days in periods of good weather. This ensures 
that most of the LOSS SNe are discovered considerably before their 
maxima. Follow-up observations of SNe are usually done starting the
night after their discovery. A detailed logging system is also
designed to keep track of the observation history of every galaxy,
which is very useful for statistical studies (e.g., SN rates).

\section*{ The LOSS discoveries in 1998 and 1999}

{\noindent  1998 discoveries:}

\hspace{2.0cm}      { Supernovae\,\,\, : 19  }

\hspace{2.0cm}      { Novae\hspace{1.1cm}\,      :  4 }

\hspace{2.0cm}       { Dwarf novae :  2 }

\hspace{2.0cm}      { Comets\hspace{0.9cm}      :  1 }

{\noindent 1999 discoveries (through mid-December):}

\hspace{2.0cm}        { Supernovae\,\,\, : 35  }

\hspace{2.0cm}        { Novae\hspace{1.1cm}\,      :  7}

\hspace{2.0cm}        { Dwarf novae    :  2}

\hspace{2.0cm}        { Comets\hspace{0.9cm}      :  1}

\vspace{10pt}

{ For a detailed list of the LOSS discoveries, please visit the 
LOSS Web page at http://astron.berkeley.edu/$\sim$bait/kait.html.}

{ Multicolor photometric observations of SNe are always emphasized 
in LOSS. Our goal is to build up a multicolor database for 
nearby SNe. So far light curves have been obtained for 11 SNe 
in 1998 and 12 SNe in 1999. Examples of the LOSS discoveries
and their light curves are presented in Figure 1. }

\bigskip
  Our supernova research at UC Berkeley is supported by NSF grant
AST-9417213 and NASA grant GO-7434.

\end{document}